\documentclass[english]{article}
\usepackage[T1]{fontenc}
\usepackage[latin9]{inputenc}

\usepackage{graphics}
\usepackage{graphicx}
\usepackage{epsfig}
\usepackage{float}
\usepackage{afterpage}

\usepackage[textwidth=120mm,textheight=200mm]{geometry}

\usepackage{units}
\usepackage{amssymb}
\usepackage{babel}

\begin{document}

\title{Learning phase transitions:comparing PCA and SVM}

\author{R. M. Woloshyn\\
TRIUMF, 4004 Wesbrook Mall\\
Vancouver, British Columbia, V6T 2A3, Canada}
\maketitle
\begin{abstract}
A comparison of results from principal component analysis and support
vector machine calculations is made for a variety of phase transitions
in two-dimensional classical spin models.
\end{abstract}

\section{Introduction}

Machine learning methods are becoming more and more prevalent in physics
research. One very active area of application is in the identification
and characterization of phase transitions. Some examples of recent
work include Refs.~[1-8]. 
An attractive feature of machine learning
techniques is that they can use the raw data from Monte Carlo simulations,
for example, the spin configurations from a spin model, to infer the
presence or nature of a phase transition without the \emph{a priori
}construction of an order parameter or other thermodynamic function. 

Most recent work focuses on the two-dimensional Ising model. However,
in Ref. \cite{Hu:2017} Hu \emph{et al. }present a more comprehensive survey
of principal component analysis (PCA) applied to Monte Carlo data
for a variety of classical spin systems. PCA is an unsupervised method
which identifies variables (principal components) whose behaviour
as a function of simulation parameters can expose the presence of
a phase transition. For example, Hu \emph{et al. }show that for the
Ising model and other phase transitions PCA identifies the magnetization
as the leading principal component without the input of any domain
knowledge. Furthermore, the pattern of principal components can be
used to distinguish crossover behaviour from a genuine phase transition.

In Ref. \cite{Ponte:2017,Giannetti:2018vif}
support vector machines (SVM) were used to study the
phase transition in the Ising model. SVM is a supervised algorithm
which can be trained using Monte Carlo data to distinguish between
the ordered and unordered phases of a spin system. The decision function
calculated for spin configurations which interpolate between the training
sets can be used as proxy for an order parameter. This was discussed
in detail for the Ising model in Ref.\cite{Giannetti:2018vif} .

The purpose of this note is
\begin{itemize}
\item to extend the SVM analysis to some spin systems which were not considered
in Ref. \cite{Giannetti:2018vif}
\item to make a comparison between the results of PCA and SVM analyses.
\end{itemize}
Sections 2 and 3 contain brief reviews of PCA and SVM respectively,
Results of analyses of Monte Carlo data for the Ising, Blume-Capel
and Biquadratic-exchange spin-1 Ising models are presented Sect. 4.
A summary is given in Sect. 5

\section{Principal component analysis}

The key feature of principal component analysis \cite{Shlens}
is the decomposition
of a data set, considered as a set of points in a large dimensional
space, along directions that are ordered by maximal variance. For
the present problem the data set will consist of an ensemble of spin
configurations on an $L\times L$ square lattice computed over a range
of a simulation parameter (temperature or a Hamiltonian model parameter).
The data can be written in the form of a matrix $\mathbf{S}$ where
each row $\mathbf{S}_{i}$ ($i=1...N)$ consists of the $n=L^{2}$
values of the spin in a configuration and N denotes the number of
rows \emph{i.e}., the number of spin configurations. The first step
is to subtract from each column its average value, call the resulting
matrix $\mathbf{X}$. Weight vectors $\mathbf{w}_{i}$($j=1...n$)
that transform $\mathbf{X}$ into principal components are defined
iteratively (see Ref. \cite{Hu:2017}) or equivalently determined from a singular
value decomposition which entails the solution of the eigenvalue problem
(see\cite{Shlens})
\begin{equation}
\label{eq:svd}
\mathbf{X}\mathbf{^{T}X\mathbf{w}_{j}=\lambda_{j}\mathbf{w}_{j}.}
\end{equation}

The final result is the principal decomposition of $\mathbf{S}$ given
by $\mathbf{P}=\mathbf{S}\mathbf{W}$ or in component form
\begin{equation}
p_{ij}=\mathbf{S}_{i}\cdot\mathbf{w}_{j}.
\end{equation}
 Hu \emph{et al. }define quantified principal components
\begin{equation}
\left\langle \left|p_{j}\right|\right\rangle =\frac{1}{N'}\sum_{i}\left|p_{ij}\right|
\end{equation}
 where the average is over a subset $N'$ of the spin configurations
which have same simulation parameter values. 

Principal component analysis is completely non-parametric and is useful
when the data have significant structure in a small number of directions,
i.e., a small number of large eigenvalues in Eqn. (\ref{eq:svd}). In the present
work the implementation of PCA in scikit-learn \cite{Pedregosa:2012toh} was used.

\section{Support vector machines}

The support vector machine is a supervised learning method that can
be used for classification and regression. In the present work it
will be used as a binary classifier. The SVM is trained to classify
elements of a data set consisting of spin configurations calculated
at two different points in the simulation parameter space. If the
training data lie on different sides of a phase transition, the ability
of the trained SVM to label spin configurations intermediate to the
training sets maybe used to investigate the phase transition, for
example, to determine the transition point in the model parameter
space.

To illustrate the concept of the SVM consider the simplest case: points
$\mathbf{x}$ in an n-dimensional space $\mathbb{R}^{n}$ with a set
of such points that can be separated into two groups (conventionally
labeled by $y=\pm$1) by a hyperplane. The points on each side of
the hyperplane that are closest to it in perpendicular distance are
the support vectors (see Sect. 1.4.7 in \cite{scikitSVM} or Fig. 2 in \cite{Giannetti:2018vif}).
The equation for a separating hyperplane takes the form%
\footnote{Of course, the symbol $\mathbf{w}$ here is not the same as the weight
vector in Sect. 2. The symbol is chosen to conform to common usage
\cite{scikitSVM}.%
} \begin{equation}
\mathbf{w}\cdot\mathbf{x}-b=0.\end{equation}
 Training the SVM involves finding $\mathbf{w}$ and $b$ which minimize
$\left\Vert \mathbf{w}\right\Vert ^{2}$ subject to $y_{i}(\mathbf{w}\cdot\mathbf{x_{i}}-b)\geq1$
for all points i in the training set. Using the solution of this minimization
problem the decision function for any point $\mathbf{x}$ is defined
to be
\begin{equation}
d(\mathbf{x})=\mathbf{w}\cdot\mathbf{x}-b.
\end{equation}
 The sign of $d(\mathbf{x})$ then assigns an arbitrary point to one
of the two groups.

In real applications a complete linear separation of the data is usually
not possible. A generalization to a so-called dual formulation can
be made which also allows for the incorporation of non-linear features \cite{Giannetti:2018vif}.
The details are omitted here and we quote only the final form of the
decision function that will be used in subsequent analyses (see \cite{Giannetti:2018vif})
\begin{equation}
d(\mathbf{x})=\frac{1}{L^{4}}\sum_{i=1}^{n_{SV}}y_{i}\alpha_{i}(\mathbf{x_{i}}\cdot\mathbf{x})^{2}+b
\end{equation}
 where $i$ labels the support vector found in training. This form
corresponds to the use of a homogeneous quadratic kernel \cite{Giannetti:2018vif}, a
kernel found to be appropriate for the Ising model and used here also
for other spin models. For this work training and calculation of the
decision function was done using functions in svm module of scikit-learn
\cite{Pedregosa:2012toh,scikitSVM}.

We will study the behaviour of spin systems as a function of a single
parameter. Suppose that the training set consists of data calculated
at parameter values%
\footnote{Although denoted by t the parameter need not be temperature. It could
be a coefficient that appears in the Hamiltonian.%
} $t_{1}$ and $t_{2}$ which may correspond to two different phases.
The decision function serves as a measure of how well the SVM can
associate data calculated for parameter values between $t_{1}$ and
$t_{2}$ with the different phases and will be used as a proxy for
an order parameter.

\section{Results}

\subsection{Ising model}

The two-dimensional Ising model is a useful testbed for methods of
studying phase transitions since the exact solution is known 
\cite{Onsager:1943jn,Yang:1952x}. The Hamiltonian is
\begin{equation}
H=-J\sum_{\left\langle i,j\right\rangle }s_{i}s_{j},
\end{equation}
 with $s_{i}=\pm$1 and the sum is over nearest-neighbour sites. For
positive $J$ there is a transition from an ordered (ferromagnetic)
spin state to an unordered (paramagnetic) state at $T_{c}/J=2/ln(1+\sqrt{2}).$
Principal component analysis of the Ising model was presented in \cite{Hu:2017}
and SVM was discussed in \cite{Ponte:2017,Giannetti:2018vif} so here we show only a few results
to set the stage for discussion of other models. 

Spin configurations were generated using a Monte Carlo method for
lattices $L$ = 64 with periodic boundary conditions at 19 values
of the temperature. Data at temperatures (in units of $J$) $\mathrm{T}{}_{1}$
and $\mathrm{T}{}_{2}$ equal 1.0 and 4.0 respectively were used for
training the SVM. Analysis was done for data at seventeen values of
the temperature spanning the range 1.5 to 2.9. At each temperature
800 well separated spin configurations were obtained.

Quantities conventionally used in the analysis of spin models, the
absolute magnetization
\begin{equation}
\left\langle \left|M\right|\right\rangle =\frac{1}{N'}\sum_{\{s\}}\left(\nicefrac{\left|\sum_{j}s_{j}\right|}{n}\right),
\end{equation}
 the squared magnetization
 \begin{equation}
\left\langle M^{2}\right\rangle ,
\end{equation}
 and the susceptibility 
 \begin{equation}
\left\langle \chi\right\rangle =n\frac{\left\langle M^{2}\right\rangle-\left\langle \left|M\right|\right\rangle ^{2}}{T}
\end{equation}
 were also computed as a function of temperature from the Monte Carlo
data to compare with what PCA and SVM learned.

\begin{figure}[phtb]

\centerline{
\includegraphics[width=120mm]{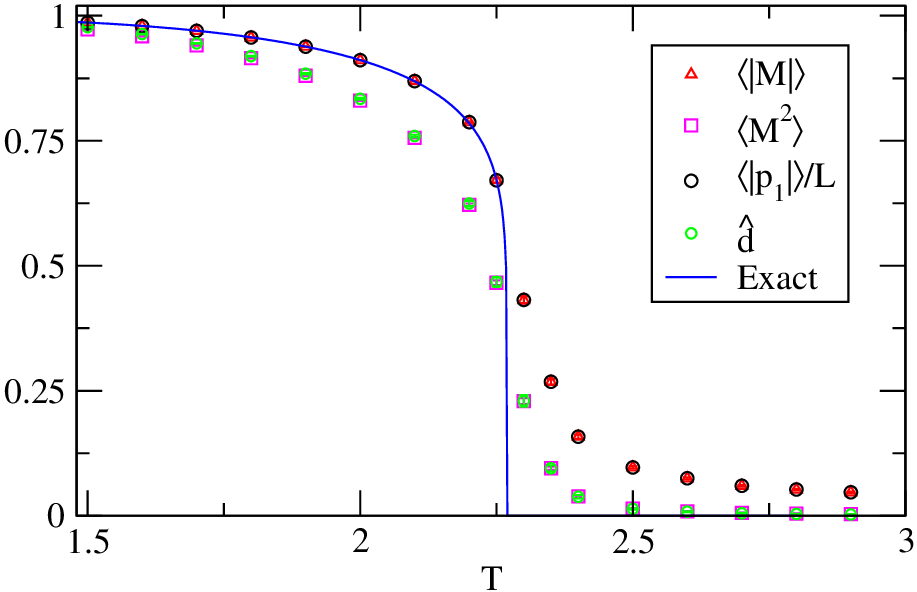}}
\caption{Summary of results for the Ising model on a lattice with $L$ = 64. The 
exact result for the magnetization \cite{Yang:1952x} is also shown (solid line). }
\label{figIsing}

\vspace*{\floatsep}


\centerline{
\includegraphics[width=120mm,height=80mm,clip=true]{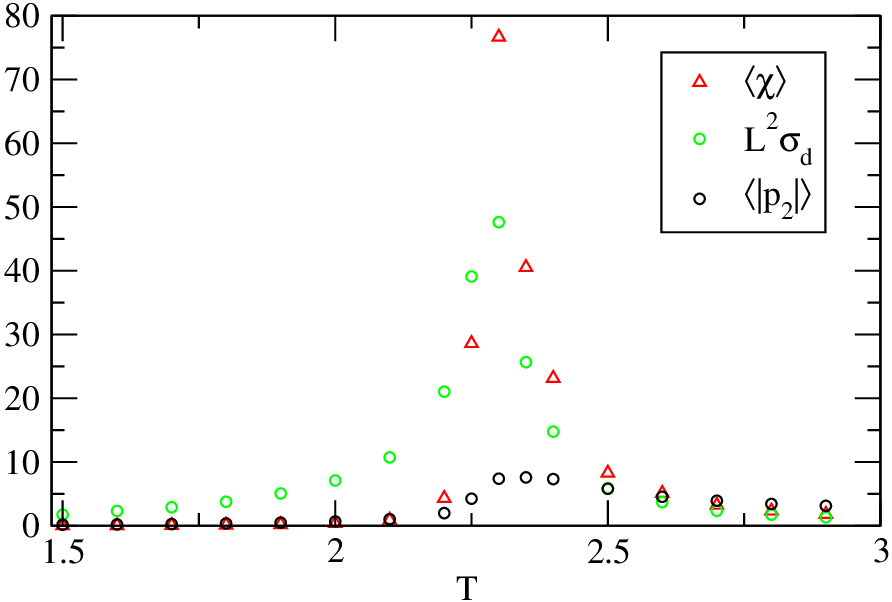}}
\caption{The scaled standard deviation of the SVM decision function 
$L^{2}\sigma_{d}$ and the quantified second principal component from PCA
$\left\langle \left|p_{2}\right|\right\rangle $ compared to the 
susceptibility for the Ising model as a function of temperature.}
\label{chiIsing}

\end{figure}

Magnetization results are summarized in Fig. \ref{figIsing}. The leading quantified
principal component $\left\langle \left|p_{1}\right|\right\rangle $
tracks $\left\langle \left|M\right|\right\rangle $ essentially exactly.
In other words, PCA picks out the magnetization as the most important
feature of the data. To illustrate what SVM learns we define a modified
decision function $\hat{d}$ by shifting and rescaling $d$ so that
$\hat{d}$ is equal to the squared magnetization at the training points
$T_{1}$ and$T_{2}$.%
\footnote{This introduces specific domain knowledge which one may not want to
do. An alternative would be to set $\hat{d}$ to span the values 1
to 0 between the training points. If training points are far from
the transition region the difference between the this procedure and
the one adopted here will be negligible.%
}
 Fig. \ref{figIsing} shows that the modified decision function reproduces the
squared magnetization. As discussed in \cite{Ponte:2017,Giannetti:2018vif} 
this is expected for
the SVM with a quadratic kernel.

In the vicinity of a continuous phase transition fluctuations on all
scales become important. This is reflected in simulations by an increase
of susceptibility (proportional to the variance of the magnetization)
in the transition region. It is suggested in \cite{Giannetti:2018vif} on the 
basis of dimensional analysis that the standard deviation of the decision 
function can also be used to identify the critical temperature. These quantities
are shown in Fig. \ref{chiIsing}. To facilitate plotting, the standard error of
$d$ (standard deviation/$\sqrt{N'})$ is used. Both quantities show
a sharp peaking in the region of the known critical temperature
\cite{Onsager:1943jn} $\approx$2.269.
As noted in \cite{Hu:2017} the second quantified principal component from
PCA $\left\langle \left|p_{2}\right|\right\rangle $ also peaks in
the region of the critical temperature. However, this feature is not
very prominent and probably not as useful for quantitative analysis
as the information from SVM.

\subsection{Blume-Capel model}

The Blume-Capel model (BCM) \cite{Blume,Capel}
is a generalization of the Ising model
which allows for three spin values $s_{i}=\pm1,0$. The Hamiltonian
takes the form
\begin{equation}
H=-J\sum_{\left\langle i,j\right\rangle }s_{i}s_{j}+\Delta\sum_{i}s_{i}^{2}.
\end{equation}
 The parameter $\Delta$ controls the density of spin 0 sites. The
transition from an ordered to an unordered spin state may be first
or second order depending on the model parameters. There is a tricritical
point at $(T/J,\Delta/J)$ = {[}0.609(4), 1.965(5){]} \cite{Xavier}. 

\begin{figure}[!htb]

\centerline{
\includegraphics[width=120mm]{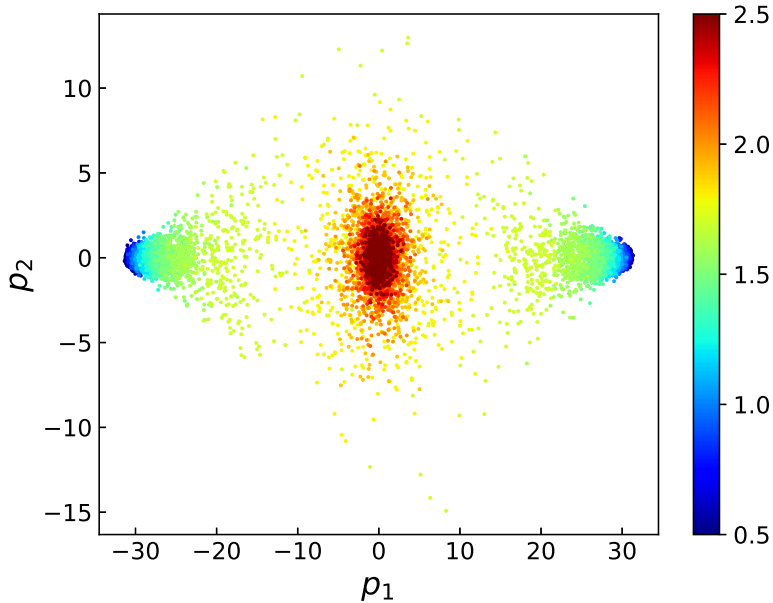}}
\caption{Scatter plot of the two leading principal components from PCA of the
Blume-Capel model at $T$ = 1, $J$ = 1 and $\Delta$ the range 0.5 to 2.5.}
\label{pcorrBCM}

\end{figure}

\begin{figure}[!htb]

\centerline{
\includegraphics[width=120mm]{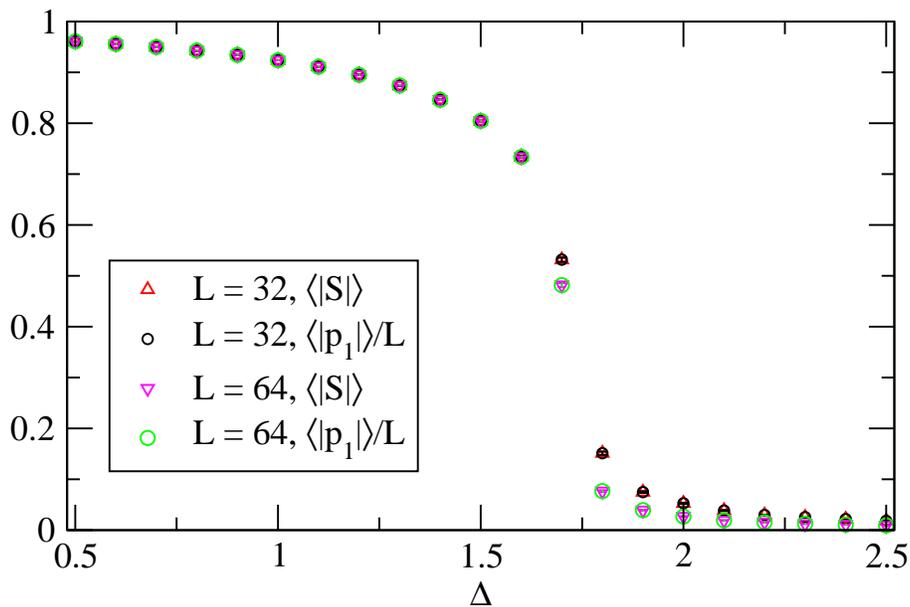}}
\caption{The quantified leading component from PCA compared to magnetization
for the Blume-Capel model at $T$ = 1, $J$ = 1 as function $\Delta$.}
\label{fig4BCM}

\end{figure}

Two different sets of model parameters were considered. The first
set has fixed $T$ = 1 and $J$ =1 with varying $\Delta.$ Monte Carlo
data were generated at $\Delta$ equal to 0.01 and 3.0 for SVM training
and in the range 0.5 to 2.5 for analysis with 1000 configurations
at each parameter point. Lattice sizes ranged from L = 32 to L = 256.

PCA was carried out for the L= 32 and L = 64 data. Fig. \ref{pcorrBCM} shows 
the scatter plot of the first two principal components for L = 32 (compare
to Fig. 5(b) in \cite{Hu:2017}). The clear separation into distinct regions
is an indicator of a transition. The presence of two lobes at larger
$|p_{1}|$ and small $p_{2}$ reflects the fact that there are
two possible ordered ground states. As in the case of the Ising model
PCA picks out the magnetization as the most prominent feature of the
data which is encoded in $p_{1}.$ This is is illustrated in Fig. \ref{fig4BCM}.

\begin{figure}[phtb]

\centerline{
\includegraphics[width=120mm]{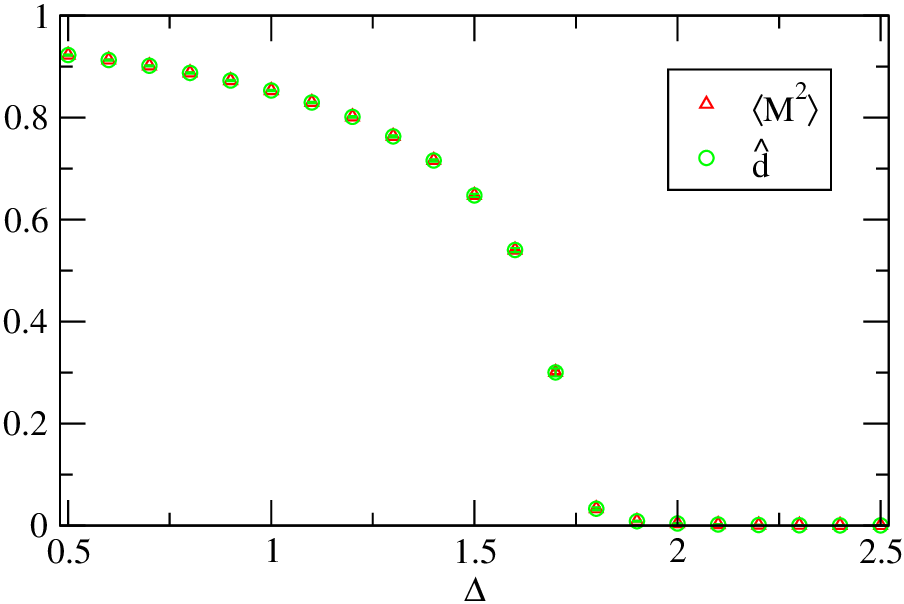}}
\caption{The modified decision function $\hat{d}$ from SVM compared to the squared magnetization
for the Blume-Capel model at $T$ = 1, $J$ = 1 as function $\Delta$ on a lattice with $L$ = 32.}
\label{figBCM32}

\vspace*{\floatsep}


\centerline{
\includegraphics[width=120mm,height=80mm,clip=true]{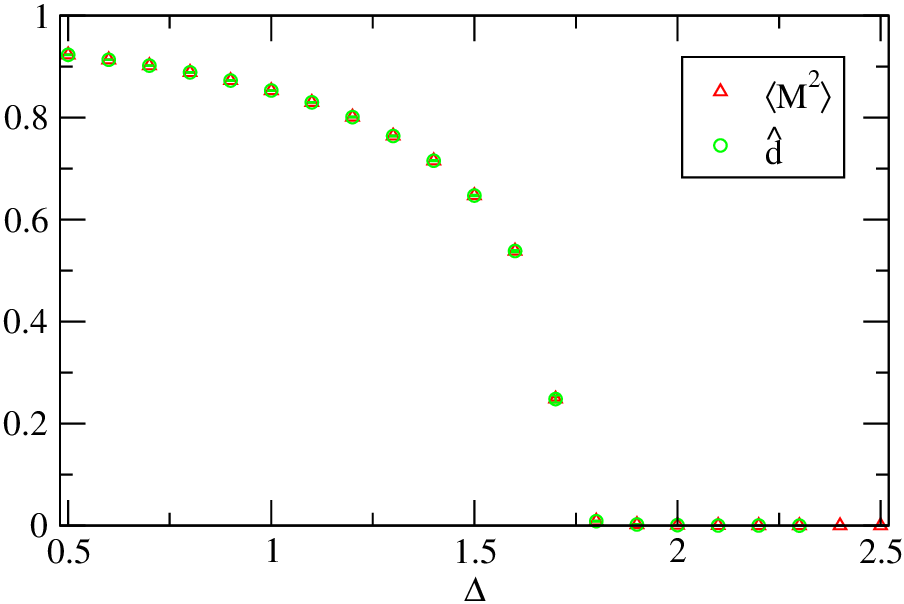}}
\caption{The modified decision function $\hat{d}$ from SVM compared to the squared magnetization
for the Blume-Capel model at $T$ = 1, $J$ = 1 as function $\Delta$ on a lattice with $L$ = 64.}
\label{figBCM64}

\end{figure}

\begin{figure}[phtb]

\centerline{
\includegraphics[width=120mm]{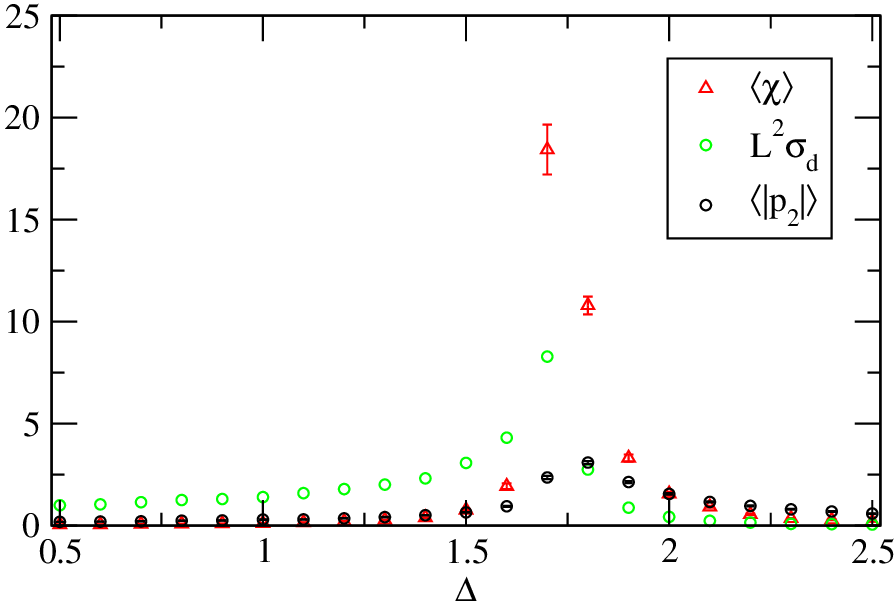}}
\caption{The scaled standard deviation of the SVM decision function 
$L^{2}\sigma_{d}$ and the quantified second principal component from PCA
$\left\langle \left|p_{2}\right|\right\rangle $ compared to the 
susceptibility for the Blume-Capel model at $T$ = 1, $J$ = 1 as a function of $\Delta$
a lattice with $L$ = 32.}
\label{figchi32}

\vspace*{\floatsep}


\centerline{
\includegraphics[width=120mm,height=80mm,clip=true]{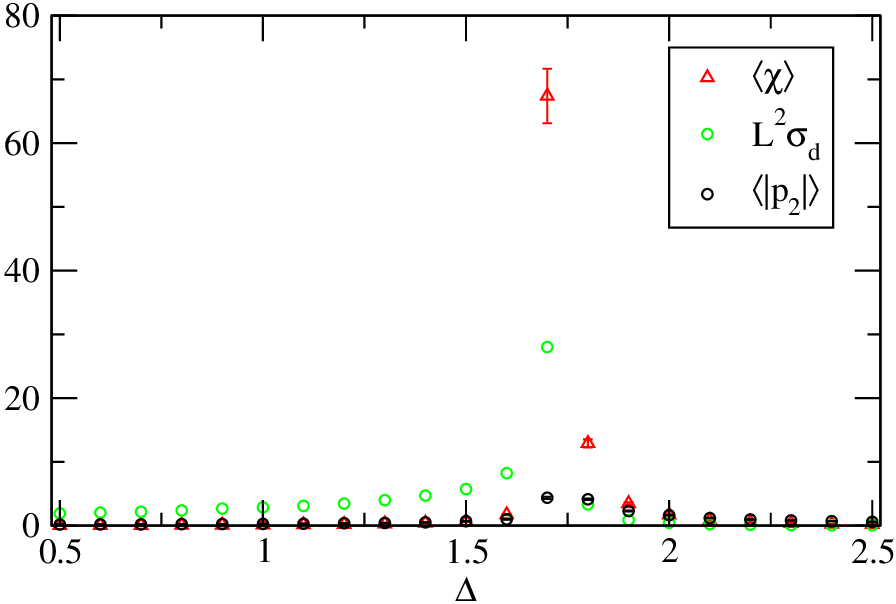}}
\caption{The scaled standard deviation of the SVM decision function 
$L^{2}\sigma_{d}$ and the quantified second principal component from PCA
$\left\langle \left|p_{2}\right|\right\rangle $ compared to the 
susceptibility for the Blume-Capel model at $T$ = 1, $J$ = 1 as a function of $\Delta$
a lattice with $L$ = 64.}
\label{figchi64}

\end{figure}

\begin{figure}[phtb]

\centerline{
\includegraphics[width=120mm]{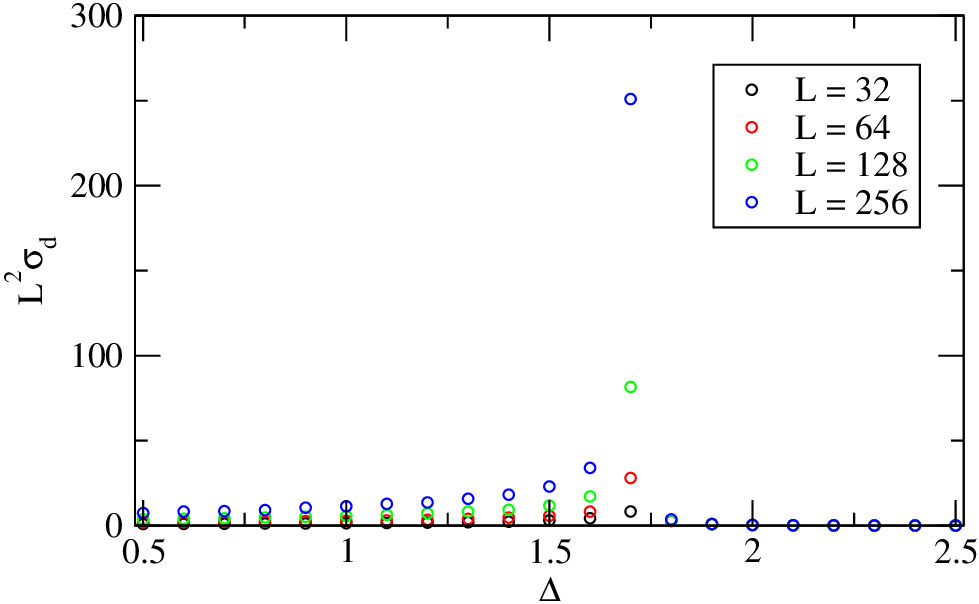}}
\caption{The scaled standard deviation of the SVM decision function 
$L^{2}\sigma_{d}$ for the Blume-Capel model at $T$ = 1, $J$ = 1 as a function of $\Delta$
on different size lattices.}
\label{figBCMsvm}

\vspace*{\floatsep}


\centerline{
\includegraphics[width=120mm,height=80mm,clip=true]{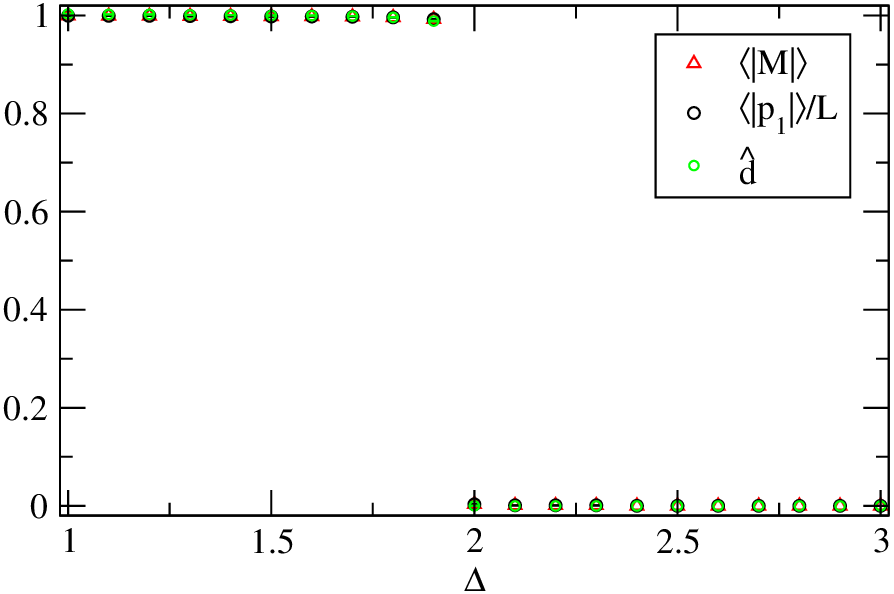}}
\caption{
The modified decision function $\hat{d}$ from SVM
and the quantified first principal component from PCA
$\left\langle \left|p_{1}\right|\right\rangle/L$ compared to the 
magnetization for the Blume-Capel model 
at $T$ = 0.4, $J$ = 1 as a function of $\Delta$ on a lattice with $L$ = 32.}
\label{figBCMt04}

\end{figure}

SVM analysis was done for lattices from L = 32 to L = 256. The modified
decision function $\hat{d}$ (pinned to squared magnetization at the
training points) is compared to the squared magnetization for L =
32 and L = 64 data in Figs. \ref{figBCM32} and \ref{figBCM64}. 
The quantified second principal
component $\left\langle \left|p_{2}\right|\right\rangle $ from PCA
and the scaled decision function standard deviation $L^{2}\sigma_{d}$
from SVM are compared to the spin susceptibility in Fig. \ref{figchi32}  and \ref{figchi64}.
The sharp peaking in the vicinity of $\triangle$ = 1.7 is consistent
with the critical $\Delta_{c}\approx$1.63 obtained in \cite{Pittman} at T
= 1. The scaled decision function standard deviation $L^{2}\sigma_{d}$
for different lattice sizes is shown in Fig. \ref{figBCMsvm}. The increased peaking
tending to a singularity as L increases is consistent with a second
order phase transition.

The second model parameter set considered has T = 0.4 and $J$ =1.
The results are shown in Fig. \ref{figBCMt04} for a lattice with L = 32. There
is a first order phase transition at $\Delta_{c}\approx$1.996 \cite{Kwak}
The outputs from PCA ($\left\langle \left|p_{1}\right|\right\rangle /L$
) and SVM ($\hat{d}$ ) correctly identify this feature.

\subsection{Biquadratic-exchange spin-1 Ising model}

The biquadratic-exchange spin-1 Ising model (BSI) is a generalization
studied in Ref. \cite{Hu:2017}. The Hamiltonian is
\begin{equation}
H=-J\sum_{\left\langle \left\langle i,j\right\rangle \right\rangle }s_{i}s_{j}+K\sum_{\left\langle i,j\right\rangle }s_{i}^{2}s_{j}^{2}
\end{equation}
 where $\left\langle \left\langle i,j\right\rangle \right\rangle $
denotes the sum over next-nearest-neighbour sites. At $J$ = 0 the
ground state is disordered with two qualitatively different spin patterns
connected by a crossover as a function of temperature. For $J\neq0$
an ordered phase appears. 

Analysis was done for two different model parameter sets. The first
set has $K$= 1, $J$ = 0. Monte Carlo data were calculated for temperatures
T from 0.08 to 2.60 on lattices from L =32 to L = 256. 
The magnetization is close to 0 at all temperatures. However,
the spin pattern is different with some short-range order appearing
at higher temperatures (see Fig. 1 in \cite{Hu:2017}). 

\begin{figure}[htb]

\centerline{
\includegraphics[width=120mm,height=80mm,clip=true]{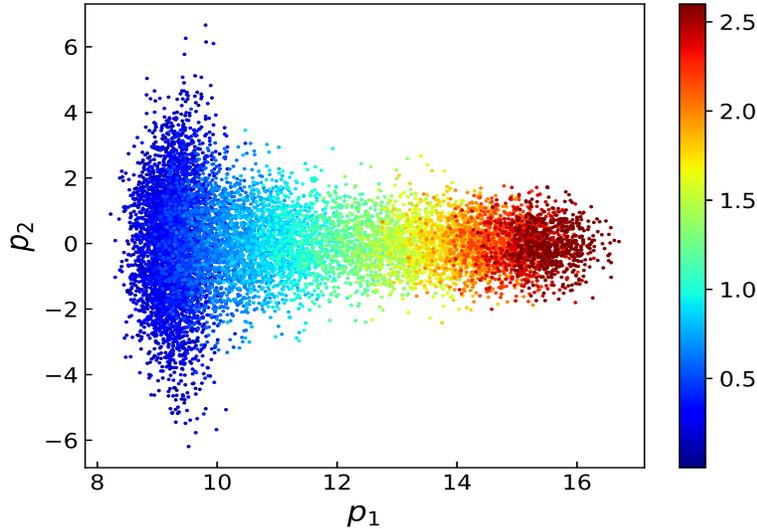}}
\caption{Scatter plot of the two leading principal components from PCA of the BSI model
at $K$ =1, $J$ = 0 and temperature range 0.08 to 2.60 with squared spin input.}
\label{figBSIpc2}

\end{figure}

\begin{figure}[htb]

\centerline{
\includegraphics[width=120mm,height=80mm,clip=true]{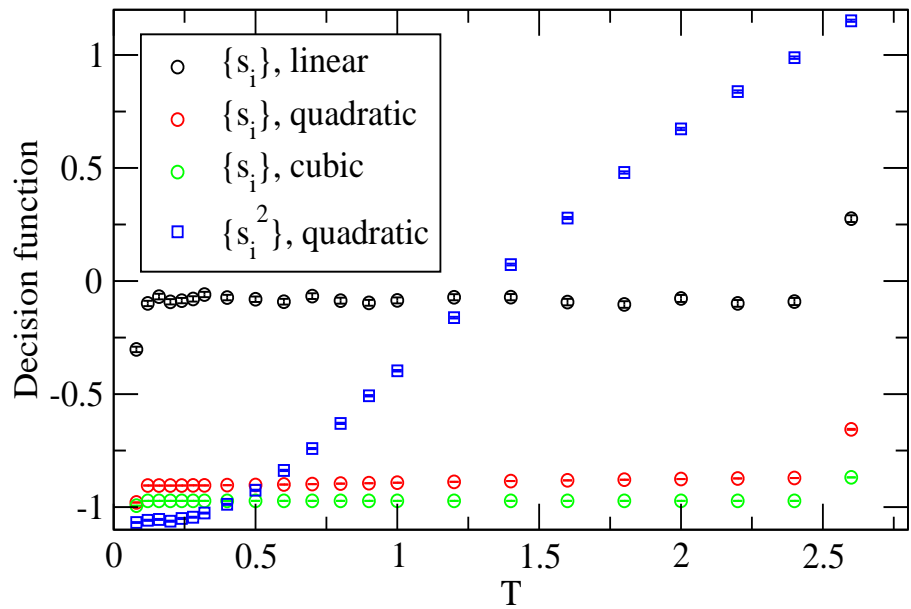}}
\caption{SVM decision function for the BSI model at $K$ =1, $J$ = 0 with different kernels
using spin input.}
\label{figBSIsvm}

\end{figure}

\begin{figure}[!ph]

\centerline{
\includegraphics[width=120mm,height=80mm,clip=true]{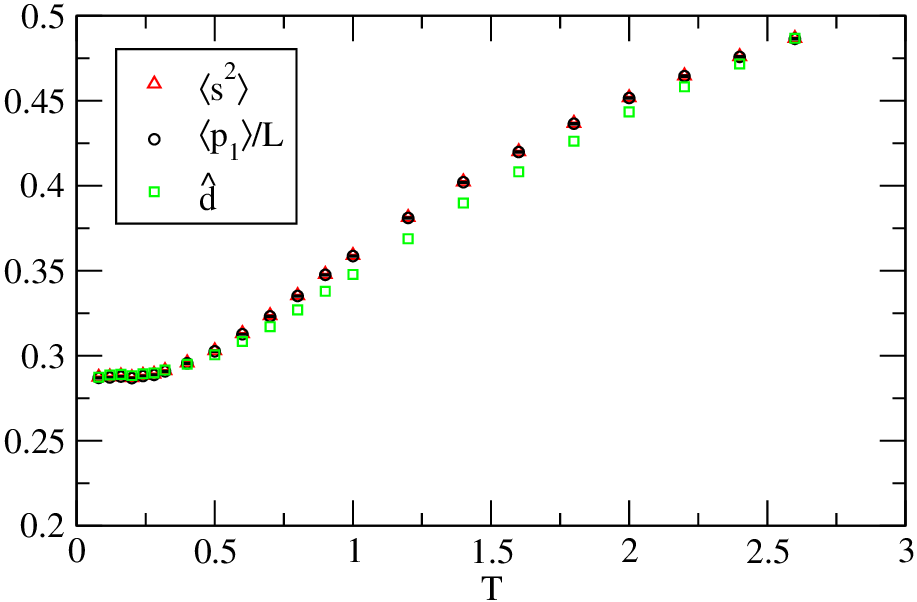}}

\caption{The modified decision function $\hat{d}$ from SVM
and the quantified first principal component from PCA
$\left\langle \left|p_{1}\right|\right\rangle/L$ compared to the expectation value of the
squared spin  $\left\langle s^{2}\right\rangle$ for the BSI model 
at $K$ = 1, $J$ = 0 as a function of $T$ on a lattice with $L$ = 32.}
\label{figBSI32}

\vspace*{\floatsep}


\centerline{
\includegraphics[width=120mm,height=80mm,clip=true]{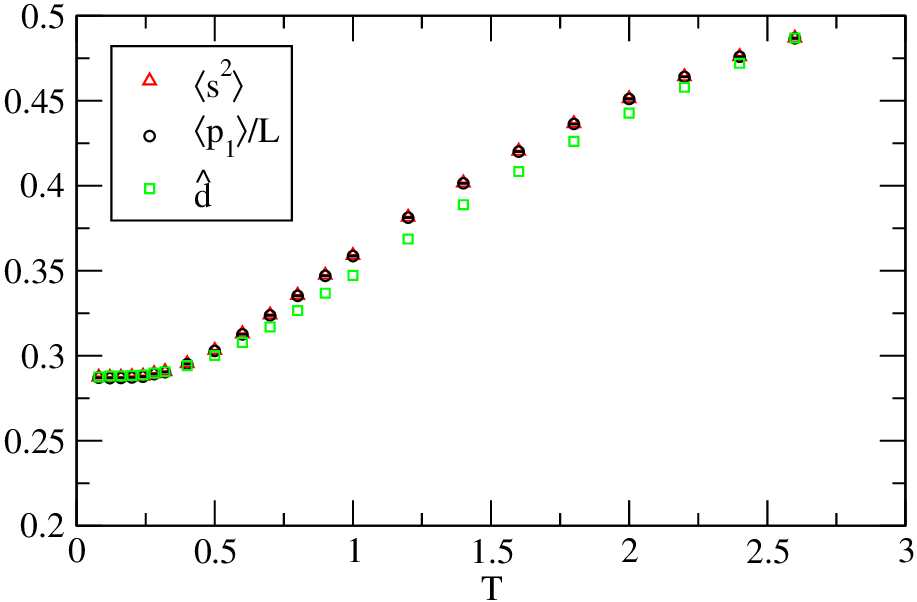}}

\caption{The modified decision function $\hat{d}$ from SVM
and the quantified first principal component from PCA
$\left\langle \left|p_{1}\right|\right\rangle/L$ compared to the expectation value of the
squared spin  $\left\langle s^{2}\right\rangle$ for the BSI model 
at $K$ = 1, $J$ = 0 as a function of $T$ on a lattice with $L$ = 64.}
\label{figBSI64}

\end{figure}

\begin{figure}[!ph]

\centerline{
\includegraphics[width=120mm,height=80mm,clip=true]{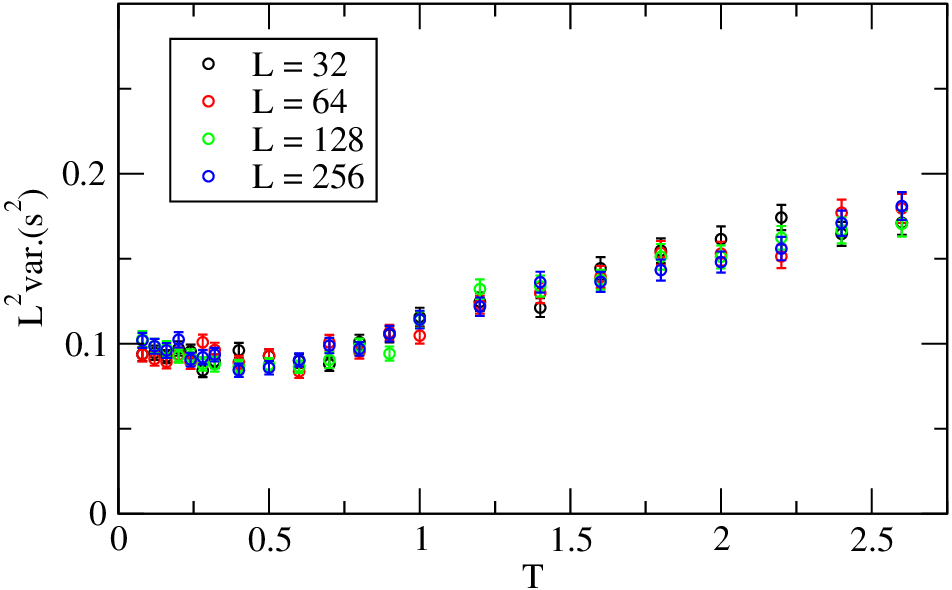}}
\caption{Variance of $s^2$ scaled by $L^2$ for the BSI model 
at $K$ = 1, $J$ = 0 as a function of $T$ on different size lattices.}
\label{figBSIvars2}

\vspace*{\floatsep}


\centerline{
\includegraphics[width=120mm,height=80mm,clip=true]{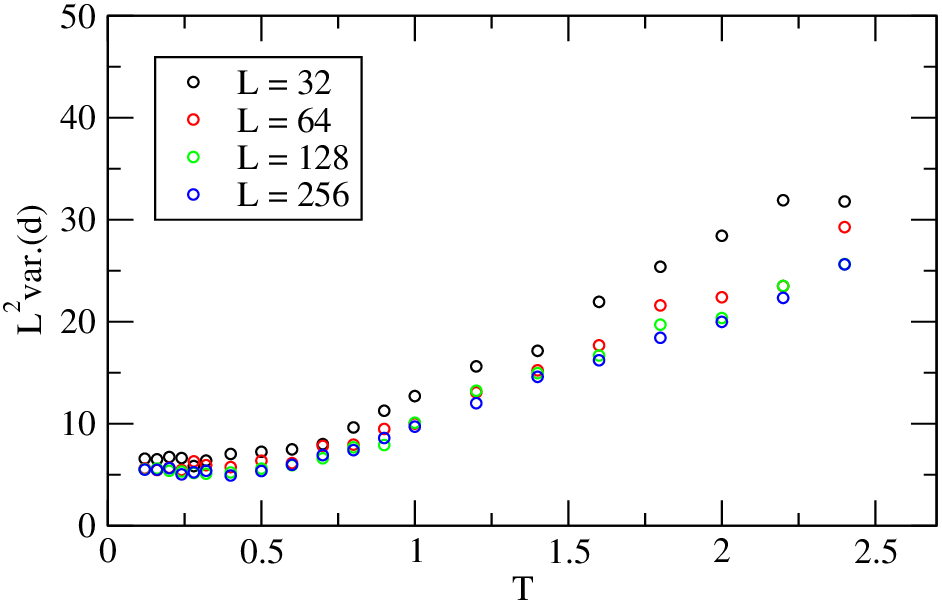}}
\caption{Variance of the SVM decision function scaled by $L^2$ for the BSI model
at $K$ = 1, $J$ = 0 as a function of $T$ on different size lattices.}
\label{figBSIvard}

\end{figure}

\afterpage{\clearpage}

As noted in \cite{Hu:2017} since PCA averages the spins over the configuration
ensemble it fails to identify any prominent feature in the spin data
and Hu \emph{et al. }suggest the use of squared spin configurations
$\{s_{i}^{2}\}$ in the analysis. Inputting these configurations gives
Fig. \ref{figBSIpc2} where one sees a dominant leading principal component which
varies smoothly with temperature. The SVM also fails to classify
correctly the spin configurations. Training with $T_{1}$= 0.08 and
$T_{2}$ = 2.60 leads to test scores well below 1. Decision functions
with spin input and different homogeneous polynomial kernels are shown
in Fig. \ref{figBSIsvm}. Also shown is the decision function with squared spin
input and a quadratic kernel. In this case the training data are classified
correctly and the decision function interpolates smoothly between
the training points.

Fig. \ref{figBSI32} and \ref{figBSI64} show the temperature dependence of the expectation
of the squared spin
\begin{equation}
\left\langle s^{2}\right\rangle =\frac{1}{N'}\sum_{\{s\}}\left(\nicefrac{\sum_{j}s_{j}^{2}}{n}\right)
\end{equation}
along with output from PCA and SVM for L= 32 and L = 64. Comparison
of these plots indicates little or no volume dependence. A question
is whether or not the change in slope of the quantities plotted in
Fig. \ref{figBSI32} and \ref{figBSI64} signals a phase transition. 
Hu \emph{et al. } \cite{Hu:2017} 
suggest that the smooth behaviour of the leading principal component and the
lack of distinct regions in Fig. \ref{figBSIpc2} (in contrast to Fig.\ref{pcorrBCM} ) indicates
a smooth crossover rather than a phase transition. A hallmark of a
continuous phase transition is increased fluctuations that extend
over the entire volume in the transition region (see Figs. \ref{chiIsing} and
\ref{figBCMsvm}). In Fig. \ref{figBSIvars2} we plot $L^{2}$ times the variance of the squared spin
as function of temperature for different volumes. No volume dependence
or sharp peaking is observed%
\footnote{Hu \emph{et al. } \cite{Hu:2017} calculate the specific heat as at function of temperature
for their Monte Carlo data(See their Fig. 7(b)). From the absence
of a significant volume dependence they suggest only a crossover behaviour.
We have repeated this calculation with the same result. %
}. Fig.  \ref{figBSIvard} shows $L^{2}$ times the variance of the decision function.
The lack of sharp peaking indicates that the SVM analysis correctly
identifies the change in the spin pattern as associated with a crossover
rather than a phase transition.

\afterpage{\clearpage}

\begin{figure}[phtb]

\centerline{
\includegraphics[width=120mm]{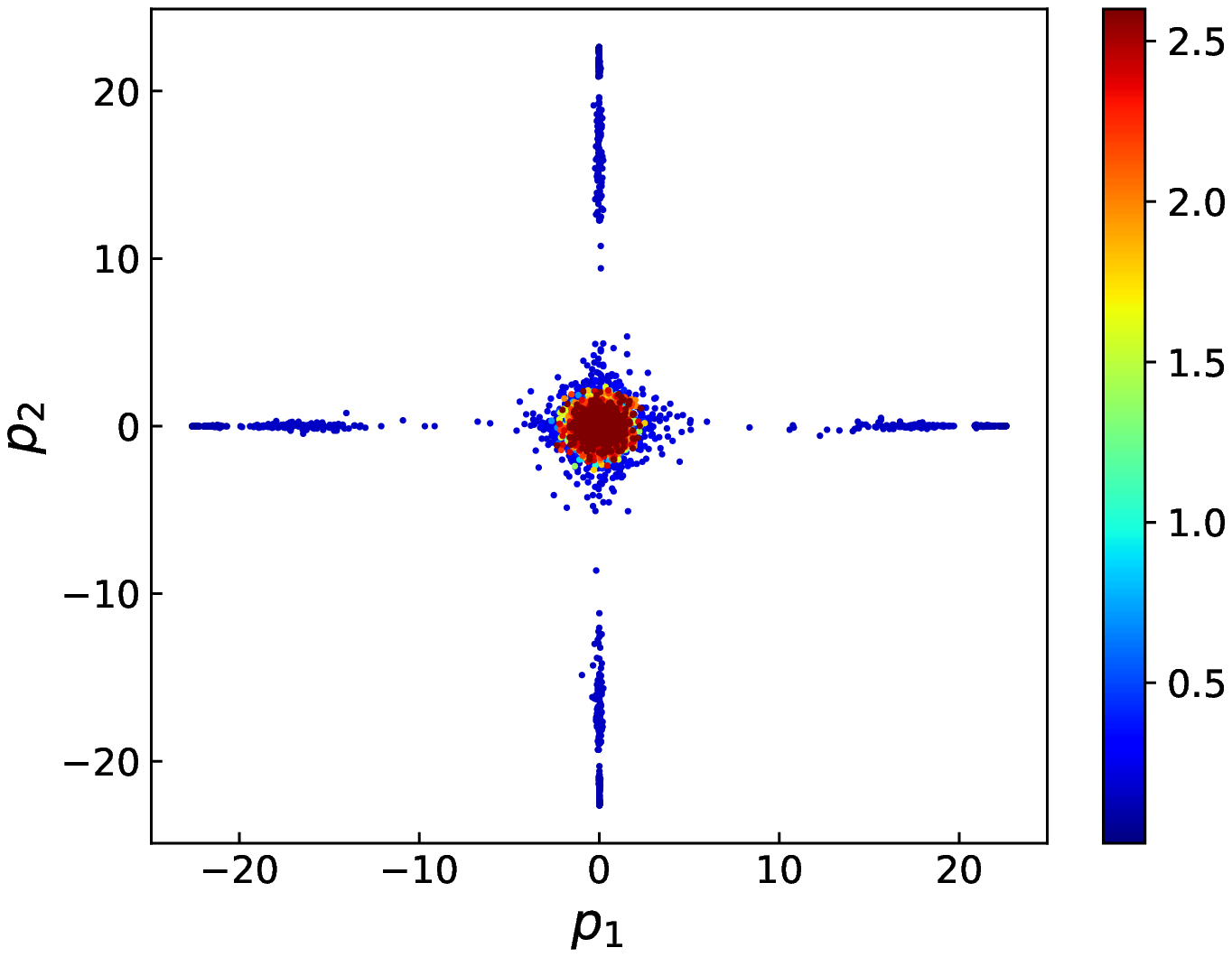}}
\caption{Scatter plot of the two leading principal components from PCA of the BSI model
at $K$ =1, $J$ = 0.1 and temperature range 0.08 to 2.60 with spin input.}
\label{figBSIpc1}

\vspace*{\floatsep}


\centerline{
\includegraphics[width=120mm,height=80mm,clip=true]{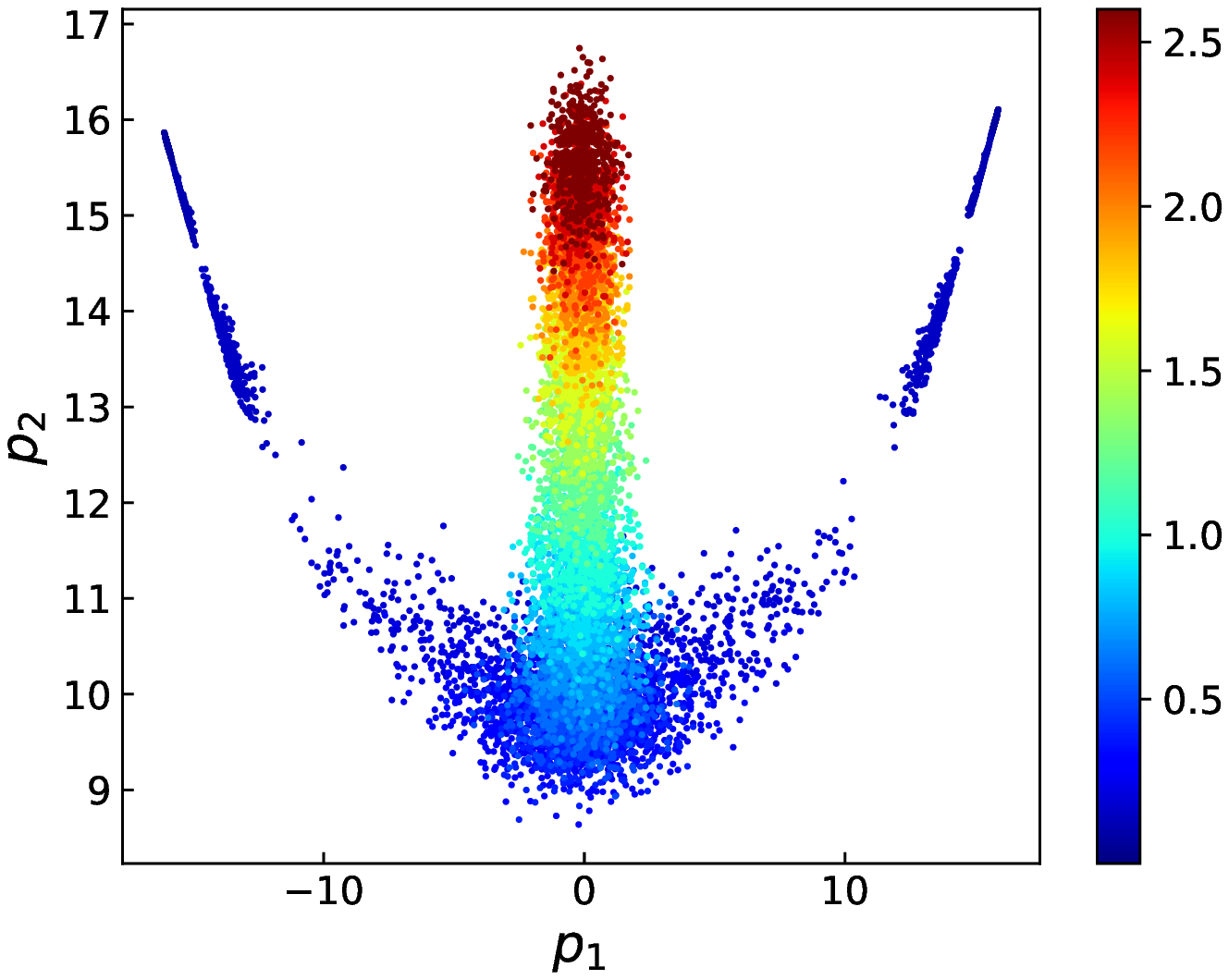}}
\caption{Scatter plot of the two leading principal components from PCA of the BSI model
at $K$ =1, $J$ = 0.1 and temperature range 0.08 to 2.60 with squared spin input.}
\label{figBSIpc21}

\end{figure}

\begin{figure}[!htb]

\centerline{
\includegraphics[width=120mm,height=80mm,clip=true]{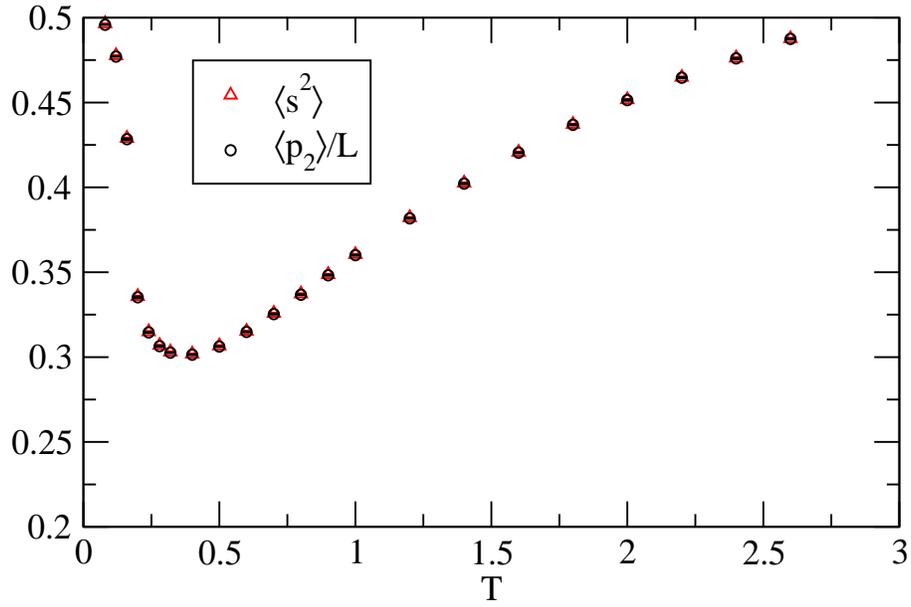}}

\caption{The quantified second principal component from PCA
$\left\langle \left|p_{2}\right|\right\rangle/L$ compared to the expectation value of the
squared spin  $\left\langle s^{2}\right\rangle$ for the BSI model 
at $K$ = 1, $J$ = 0.1 as a function of $T$ on a lattice with $L$ = 32.}
\label{figBSIp2}

\end{figure}

The second set of model parameters has $K$ = 1, $J$ = 0.1.  At low temperatures an ordered
phase with a checker board pattern of zero and nonzero spins appears (see Fig. 1
in \cite{Hu:2017}).
PCA was carried out for spin simulation data in the temperature range
0.08 to 2.60. The scatter plot of the two leading principal components
is shown in Fig. \ref{figBSIpc1}. The four arms show the presence of four possible
ordered ground states at low temperature. Using the squared spin configurations
$\{s_{i}^{2}\}$ as input the resulting $p_{1}$ versus $p_{2}$ correlation
is given in Fig. \ref{figBSIpc21}. The interpretation of the principal components
is not as obvious here as it was for the Ising or Blume-Capel model.
Fig. \ref{figBSIp2} shows that with squared spin input the second quantified principal
component reproduces the expectation value $\left\langle s^{2}\right\rangle .$

\begin{figure}[!ph]

\centerline{
\includegraphics[width=120mm,height=80mm,clip=true]{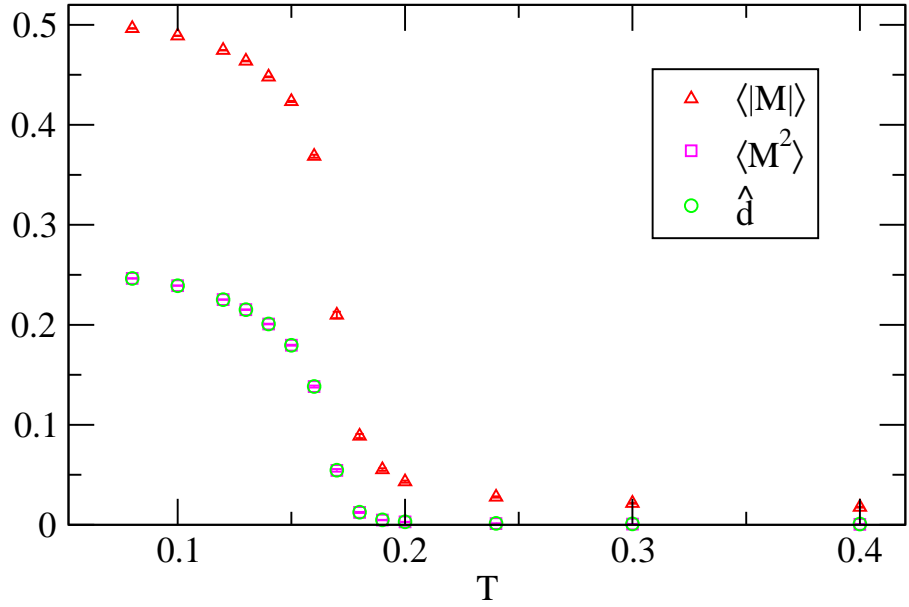}}

\caption{The modified decision function $\hat{d}$ from SVM
compared to magnetization and squared magnetization for the BSI model 
at $K$ = 1, $J$ = 0.1 as a function of $T$ on a lattice with $L$ = 32.}
\label{figBSIdfhat}

\end{figure}
\begin{figure}[!htb]

\centerline{
\includegraphics[width=120mm,height=80mm,clip=true]{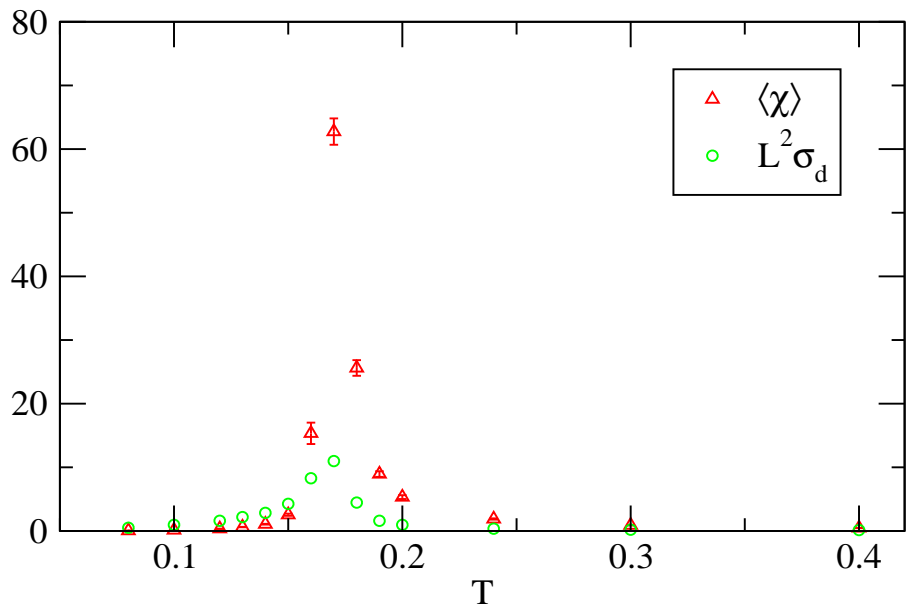}}

\caption{The scaled standard deviation of the SVM decision function 
$L^{2}\sigma_{d}$ compared to the 
susceptibility for the BSI model at $K$ = 1, $J$ = 0.1 as a function of $T$
on a lattice with $L$ = 32.}
\label{figBSIchiDF}

\end{figure}

\begin{figure}[!ph]

\centerline{
\includegraphics[width=120mm,height=80mm,clip=true]{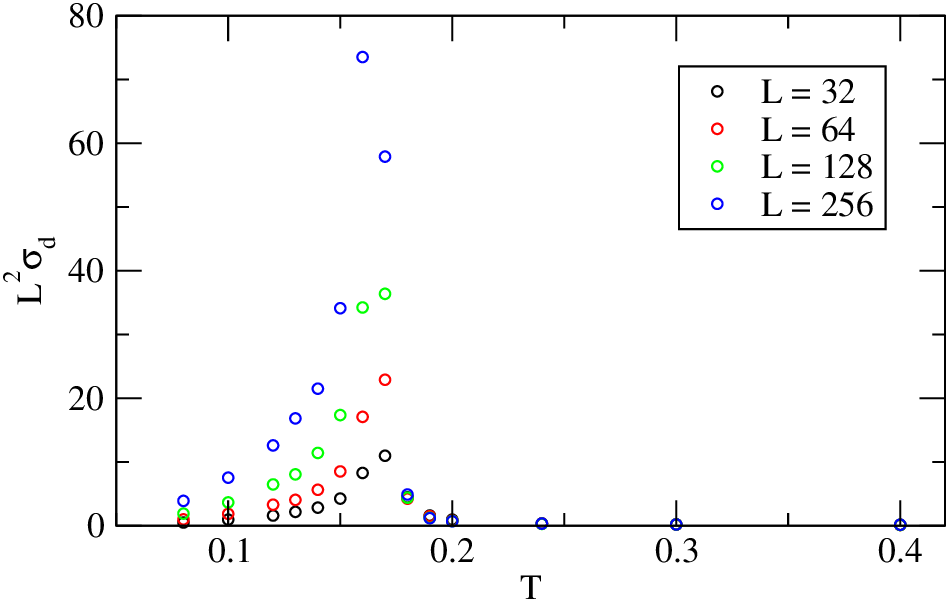}}

\caption{The scaled standard deviation of the SVM decision function 
$L^{2}\sigma_{d}$ for the BSI model at $K$ = 1, $J$ = 0.1 as a function of $T$
on different size lattices.}
\label{figBSIDFvol}

\vspace*{\floatsep}

\centerline{
\includegraphics[width=120mm,height=80mm,clip=true]{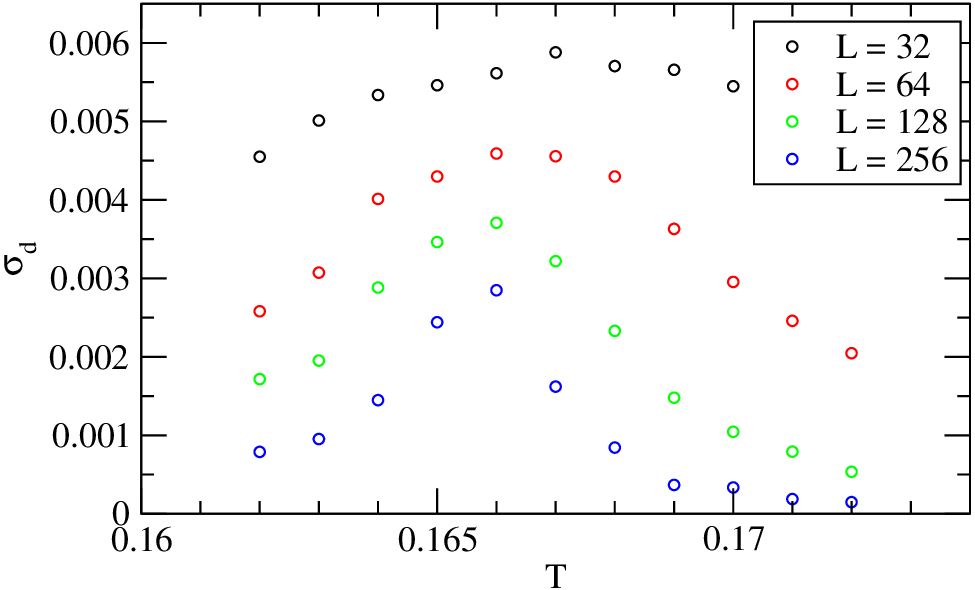}}
\caption{Standard deviation of the decision function versus temperature in the transition
region for the BSI model at $K$ = 1, $J$ = 0.1 on different size lattices..}
\label{figBSIsigma}

\end{figure}

SVM analysis was done using spin configuration input with $T_{1}$=
0.08 and $T_{2}$ = 0.60 for training. The results for the modified
decision function along with magnetization and squared magnetization
are plotted in Fig. \ref{figBSIdfhat} for a lattice with L = 32. As with other models
the modified decision function tracks the squared magnetization. The
scaled standard deviation of the decision function as function of
temperature is compared with the susceptibility $\left\langle \chi\right\rangle $
in Fig. \ref{figBSIchiDF}. The peaking is indicative of a phase transition in the
region $T$ = 1.7. To verify that there is a phase transition and
to do a quantitative analysis volume dependence has to be examined.
Fig. \ref{figBSIDFvol} shows the lattice size
dependence of $L^{2}\sigma_{d}$ for lattices up to $L$ = 256. 

\begin{figure}[!htb]

\centerline{
\includegraphics[width=120mm,height=80mm,clip=true]{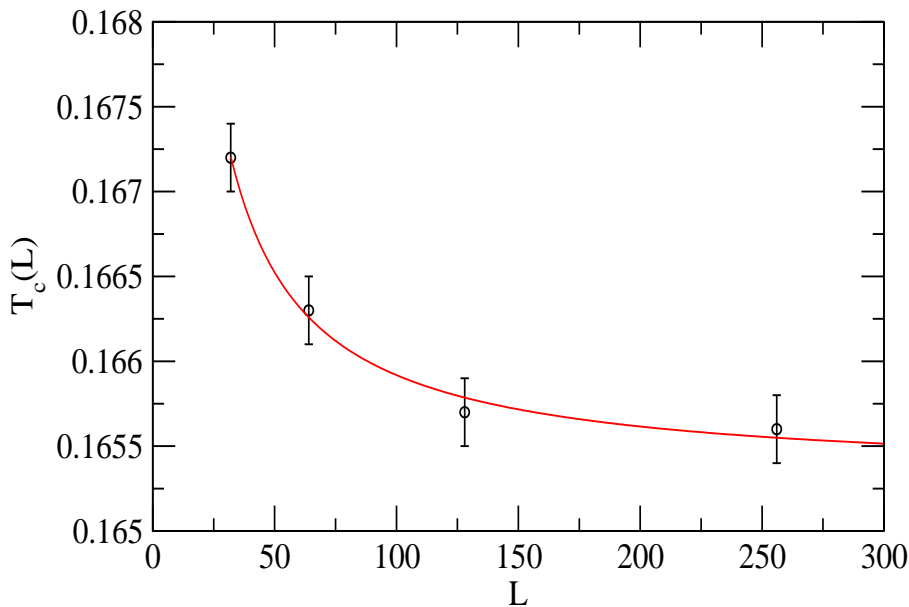}}
\caption{Estimated peak values $T_{c}(L)$ versus $L$ along with an
extrapolation $T_{c}+c/L$ (solid line).}
\label{figBSITc}

\end{figure}

To determine the critical temperature we use $L^{2}\sigma_{d}$ as
proxy for the susceptibility and estimated the temperature of the
peak position of this quantity for different volumes. Call this $T_{c}(L).$
Then $T_{c}(L)$ extrapolated to infinite volume gives $T_{c}.$ To
get the peak position Monte Carlo data with 4000 configurations per
temperature were generated for a fine scan with 0.162$\leq T\leq$0.172.
The error in the decision function trained with $T_{1}$= 0.08 and
$T_{2}$ = 0.60 is plotted in Fig. \ref{figBSIsigma}. The estimated 
peak values $T_{c}(L)$
along with an extrapolation $T_{c}+c/L$ are shown in Fig. \ref{figBSITc}. The
inferred critical temperature is 0.1655(5), slightly larger than that
value 0.163 given in \cite{Hu:2017}.

\section{Summary }

Machine learning methods provide the opportunity to study phase transitions without 
\emph{a priori} specification of an order parameter or other thermodynamic function.
In this work two different methods, principal component analysis and support vector 
machines, have been applied to the Blume-Capel and Biquadratic Spin-1 Ising models.
These classical spin models can exhibit different transition behaviour for different
values of their model parameters. PCA and SVM can expose both the presence and nature of
these transitions.

PCA is an unsupervised method which is useful when the data have a few dominant features.
As discussed in \cite{Hu:2017}, and reproduced here, a plot of the leading principal components
provides useful qualitative information about the model's phase behaviour using as input
raw Monte Carlo simulation data (see Figs. 3,13 ). PCA typically identifies the 
magnetization as the dominant feature without input of any domain knowledge.

SVM is a supervised method trained for the models studied here as a binary classifier
for Monte Carlo data generated at two different points in temperature or in the model
parameter space. The trained decision function can be used to determine the presence of
a phase transition separating training points. We showed examples from the  
Blume-Capel and Biquadratic Spin-1 Ising models how the standard deviation of the 
decision function interpolated between the training points provides a proxy for the 
susceptibility. As suggested in \cite{Giannetti:2018vif}, SVM can provide quantitative information without 
\emph{a priori} knowledge of an order parameter. As an example of a quantitative
analysis the critical temperature for the BSI model at $K$ = 1, $J$ = 0.1 was determined
to be 0.1655(5), close to the value given in \cite{Hu:2017}.

\section*{Acknowledgment}

TRIUMF receives federal funding via a contribution agreement with
the National Research Council of Canada.


\end{document}